\documentclass[12pt]{iopart}

\usepackage{iopams}
\usepackage{color}
\usepackage{graphicx}
\usepackage{dcolumn}
\usepackage{bm}

\usepackage[utf8]{inputenc}
\usepackage[T1]{fontenc}
\usepackage{mathptmx}
\usepackage{enumerate}
\usepackage{theorem}
\usepackage{xcolor}

\usepackage{caption}
\usepackage{subcaption}

\newtheorem{thm}{Theorem}

\newtheorem{prp}{Proposition}
\newtheorem{cor}{Corollary}
\newtheorem{defi}{Definition}

\newcommand*{\num}[1]{\scalebox{0.7}[0.8]{$#1$}}

\newcommand{\proof}{\noindent {\bf Proof:} \hspace{0.1in}}
\newcommand{\qed}{\hfill\mbox{\raggedright $\Box$}\medskip}

\newcommand{\smtimes}{\,{\scriptstyle \times}\,}

\newcommand{\smotimes}{\,{\scriptstyle \otimes}\,}

\newcommand{\p}{\partial}
\newcommand{\R}{\mathbb{R}}
\newcommand{\M}{\mathcal{M}}

\newcommand{\Diff}{\mathcal{C}^\infty}

\newcommand{\submetric}{\ensuremath{\mathbf{h}}}
\newcommand{\metric}{\ensuremath{\mathbf{g}}}
\newcommand{\normal}{\ensuremath{\mathbf{u}}}

\newcommand{\hubble}{\ensuremath{\mathrm{H}}}
\newcommand{\eqstate}{\ensuremath{\gamma}}
\newcommand{\cell}{\ensuremath{\mathrm{K}}}
\newcommand{\laplacian}{\,\raisebox{0.01em}{\scalebox{1.2}[1.3]{${\scriptstyle \Delta}$}}}

\begin{document}

\title{On the intrinsically flat cosmological models in a lattice}

\author{Eduardo Bittencourt, Leandro G. Gomes, Grasiele B. Santos}
\address{Federal University of Itajub\'a, Av. BPS, 1303, Itajub\'a-MG, 37500-903, Brazil}
\eads{\mailto{lggomes@unifei.edu.br}, \mailto{bittencourt@unifei.edu.br}, \mailto{gbsantos@unifei.edu.br}}

\date{\today}

\begin{abstract}

In this manuscript we investigate the intrinsically flat (space-flat) spacetimes as viable cosmological models. We show that they have a natural geometric structure which is suitable to describe inhomogeneous matter distributions forming a periodic pattern throughout the space. We prove theorems for their local representation and for existence and uniqueness of the Einstein's equations with these periodic boundary conditions. We also find an interesting class of exact solutions, which illustrates the applicability of such spacetimes in cosmology, with an early time behavior close to homogeneity and isotropy and a late time aspect with peaks and voids in the matter distribution.
\end{abstract}

\maketitle

\section{\label{sec:Intro} Introduction \protect}

The concepts of group action and symmetry play a central role in the theory of general relativity, for they are guidelines for probing the nonlinear Einstein's equations. These ideas are behind the mathematical description of one of the most fundamental tenets in modern cosmology, known as the cosmological principle: the universe is spatially homogeneous and isotropic on the large-scale average (see \cite{Peebles} for a historical account of it). This context of the standard cosmological model is mathematically characterized by the presence of the maximal degree of spatial symmetry in the spacetime $(\M,\metric)$ \cite{Wald}, or at least its local version, which is locally equivalent, but richer in the possibilities for the choice of the global topology \cite{Ring}. These are the Robertson-Walker (RW) spacetimes.

Some decades ago, Collins and Szafron \cite{coll_79,coll1_79,coll2_79} called the attention to the cosmological applications of a more general concept, called \textit{intrinsic symmetry}. In a suitable context for our purposes here, let us take a splitting into time plus space of the manifold $\M$, i.e. $\mathcal{M}=\R \times \Sigma$, denote each space section by $\Sigma_t:=\{t\}\times \Sigma$ and the restricted (Riemannian) space metric by $\submetric(t):=\metric_{|_{\Sigma_t}}$, with $\metric$ as the spacetime metric. A diffeomorphism $f: \M \to \M$ is an intrinsic symmetry of this structure if it preserves the leaves of the foliation, $f(\Sigma_t)=\Sigma_t$, and is an isometry of $\submetric(t)$, for every $t$. Note that $f$ does not need to be an isometry of the spacetime metric $\metric$. Therefore $\M$ could be ``space-homogeneous'' , in the sense that each leaf is a homogeneous Riemannian space, without being spatially homogeneous, where the intrinsic symmetries turn out as spacetime isometries, just like the RW and Bianchi models \cite{ellis_mac_marteens}.

In this paper we scrutinize the class of the space-flat spacetimes, which are not necessarily spatially flat, and probe its potentiality as cosmological models. As we shall see later, the theorems demonstrated along the text ensure that such class encompass the following interesting properties:
\begin{itemize}
\item For physically motivated equations of state, the Einstein's equations simplify enough in order to provide existence and uniqueness theorems using known results from the theory of nonlinear partial differential equations. Furthermore, the initial conditions can be chosen closely to the inhomogeneous spatial distribution of matter observed today \cite{abbott2022dark} (Theorem \ref{Thm:ExistenceUniqueness}).
\item The time evolution is left apart as we use the scale factor as a time variable. As a consequence, the Hubble parameter is a free function which is able to fit, a priori, any late time observations of luminosity distances and redshift relation, with no need of a cosmological constant (section \ref{Sec:GeneralAspects_EinsteinEquations}).
\item The spatial distribution of matter forms a lattice in which the fundamental domain (cosmological cell) is filled with an inhomogeneous energy density under periodic boundary conditions.
\item The solutions are consistent with the standard viewpoint of the Universe emerging from a highly uniform early phase, and as it expands, it becomes less homogeneous (See Corollary \ref{Cor:HomogeneousSmallerScaleFactor}).
\end{itemize}

The space-flat cosmological models should also be put in the class of the spacetimes based on their less symmetrical structure, as compared to the RW's ones, offer different perspectives in describing the observed universe. In particular, the prospect of using them to give a realistic alternative to dark energy is feasible \cite{BGS}. In this vein, there has been a number of attempts to explain the apparent accelerated expansion in the standard model as a result of inhomogeneities. One of the most accepted alternatives to dark energy concerns the inhomogeneous Lema\^itre-Tolman-Bondi models, claiming that the accelerated expansion is a fictitious effect due to the fact that we are located close to the center of a void \cite{clifton08,feb10}. Also, backreaction terms due to small scale inhomogeneities could mimic an accelerated expansion on larger scales \cite{kolb,buchert,wiltshire}. Besides that, there is the idea of constructing relativistic cosmological models with a discrete matter distribution, which is not a new one. It can be achieved, for example, as using black-hole lattices (see, for instance, reference \cite{bentivegna}). 

The paper is organized as follows: in the section \ref{Sec:GeneralAspects} we obtain the first results on space-flat Lorentzian manifolds, such as the theorem for their local representation and the form that the Einstein's equations take in the adapted coordinates, as well as we set the boundary conditions suitable for the cosmological applications. The section \ref{Sec:MainTheorems} is devoted to the main theorem on the existence and uniqueness of solutions of Einstein's equations with the equation of state $p=(\eqstate (t)-1)\rho$. As a consequence, we also prove an ``early time'' (small scale factor) behavior close to the RW model. In the final section, we present the concluding remarks. We followed the reference \cite{ellis_mac_marteens} for the conventions of the spacetime indexes, the signature of the metric and the definition of tensors in general.

\section{The General Setting}\label{Sec:GeneralAspects}

\subsection{General aspects of a space-flat spacetime}

Let $\normal$ be a unitary time-like vector field ($\normal^2=-1$) in a $m$-dimensional spacetime $(\M,\metric)$ and denote its dual $1$-form as $\normal^\flat = \metric(\normal, \cdot)$, that is, $u^\flat_\nu=u_\nu$ is obtained by lowering the index of $u^\nu$. Hence, the orthogonal distribution $\normal^\bot$ is just the kernel $\ker \normal^\flat$, given by the vectors orthogonal to $\normal$ at each point in $\M$. We recall that $\normal$ is vorticity-free \cite{ellis_mac_marteens} if for all $\mathbf{X},\mathbf{Y} \in \normal^\bot$ we have $\metric (\nabla_\mathbf{X}\normal, \mathbf{Y}) = \metric (\nabla_\mathbf{Y}\normal, \mathbf{X})$, implying that $\metric (\nabla_\mathbf{X}\mathbf{Y}, \normal )$ is symmetric. Since the Levi-Civita connection is torsion-free, this is equivalent as demanding the orthogonal distribution to be integrable, a property we will assume throughout the text. Each maximal integral submanifold passing through $p\in \M$ is a Riemannian manifold with the metric $\submetric = \metric + \normal^\flat \smotimes\normal^\flat$. It is called the space section at $p$ and denoted as $\Sigma_p$.
\begin{defi}
\label{def_charac_u}
A spacetime $(\M,\metric)$ is said to be \emph{space-flat}, or \emph{intrinsically flat}, if there is a timelike, vorticity-free, vector field $\normal$ such that their space sections with the inherited geometry are flat Riemannian manifolds.
\end{defi}

Our first proposition is concerned with the local representation of the space-flat spacetimes.

\begin{prp}\label{Thm:LocalFormFlatSpaceMetric}
In the vicinity of any point in a space-flat spacetime $(\M,\metric)$, with  $\normal$ the vector field as in the definition \ref{def_charac_u}, there are coordinates $(x^\mu)=(t,x^i)$, with $i=1,\dots,m-1$,  for which  $\normal = e^{-\phi(x^{\mu})} \,\frac{\partial}{\partial t}$ and the components of the space metric $\submetric$ are all time functions:
\begin{equation}\label{Eq:Metric_General_Flat}
 \metric = - \, e^{2 \phi}\, dt\otimes dt +  h_{ij}(t) \, dx^i \otimes dx^j \, .
\end{equation}
In particular, $\normal$ is shear-free if, and only if,  this formula can be reduced to
\begin{equation}\label{Eq:Metric_Shearfree}
 \metric = - \, e^{2 \phi}\, dt\otimes dt + a(t)^2 \, \delta_{ij} \, dx^i\otimes dx^j \, .
\end{equation}
In both cases, we can set $\phi=0$ without loss of generality if, and only if, $\normal$ is geodesic, that is, $\nabla_\normal\normal=0$.
\end{prp}
\proof
Since $\normal$ is vorticity-free, there are coordinates adapted to the splitting of space + time generated by it with \cite{ellis_mac_marteens}
\[
\normal = e^{-\phi } \,\frac{\partial}{\partial t}
\quad \textrm{and} \quad
\metric = - \, e^{2 \phi}\, dt\otimes dt +  \tilde{h}_{ij}(t,y) \, dy^i \otimes dy^j \, ,
\]
where the flat space section $\Sigma_{t_0}$ passing through the point $(t_0,y_0)$ is locally characterized by $t=t_0$. Denoting the spatial projection of the gradient of $\phi$ by $\bar\nabla \phi$, $\phi$ is a function that can be set to zero without loss of generality if, and only if,  it is a function of $t$ only, that is, $\nabla_{\normal} \normal=\bar\nabla \phi= 0$.

The intrinsically flat coordinate system is built in the following way: Let $F_0^t$ denote the flux of the vector field $\partial_t$ and fix $p \in \M$ at the space section $t=0$ with coordinates $(0,y_0)$.  Define the spatial frame along the curve $F^t_0(0,y_0)=(t,y_0)$ as
\[
\mathbf{e}_1(t,y_0):= \frac{\p}{\p y^1} \quad, \,  \ldots
\mathbf{e}_{m-1}(t,y_0):= \frac{\p}{\p y^{m-1}} \, .
\]
Along each flat hypersurface $t=t_0$ we can locally propagate this frame such that the $\mathbf{e}_i$'s are all Killing fields of $\submetric$ representing local translations, that is, $[\mathbf{e}_i,\mathbf{e}_j]=0$. As we take $F_i^s(p)$ to represent the flux of $\mathbf{e}_i$, the map
\begin{equation}
\psi(t,x^1, \ldots , x^{m-1}):= F_{1}^{x^1} \circ \ldots \circ F_{m-1}^{x^{m-1}}\, (t,y_0)     
\end{equation}
gives us the desired coordinate transformation. In fact, $\psi$ leaves any space section invariant, since the flux of each field $\mathbf{e}_k$ do it. Furthermore, as they all commute with each other, the fluxes also commute, so that  $\mathbf{e}_k$ is represented by $\partial/\partial x^k$ in the new coordinates. But they are Killing fields of $h_{ij}(t,x) \, dx^i \otimes dx^j$ for each $t$, that is, $\partial h_{ij}/\partial x^k=0$. \qed

In the dimension $m=4$, the theorem  \ref{Thm:LocalFormFlatSpaceMetric} states that any space-flat spacetime is locally a Bianchi-I model if, and only if, $\normal$ is geodesic. In particular, the flat RW case follows as we take $\normal$ to be both geodesic and shear-free. This shows that space-flatness allows suitable generalizations of the models often considered in Cosmology by including both inhomogeneity, as $\phi$ depends on the space coordinates, and anisotropy, presented in the shear tensor of $h_{ij}(t)$. In this manuscript, we shall focus mainly in the inhomogeneity property, so that in the most parts of it we will assume $\normal$ to be shear-free. 

There are non-orthogonal versions of the proposition \ref{Thm:LocalFormFlatSpaceMetric} in the literature (see \cite{coll2_79}, section 2). We shall emphasize that our results are valid in those contexts as well as in any spacetime which is foliated by conformally flat spatial sections. To see that, it is enough to take $\normal$ to be the vector field normal to the conformally flat foliation, which is clearly vorticity-free, and apply the proposition  \ref{Thm:LocalFormFlatSpaceMetric} to $e^{2\xi}\, \metric$, for an appropriate function $\xi$. Since the theorem is local, this function need not to be globally defined. Therefore, that result can be readily extended to any spacetime foliated by conformally flat slices:
\begin{cor}\label{Cor:LocalFormFlatSpaceMetric}
If $(\M,\metric)$ is foliated by conformally flat spatial sections, then around each point there are coordinates where the metric is represented as
\begin{equation}\label{Eq:Metric_General_Conformally_Flat}
 \metric = - \, e^{2 \phi(t,x)}\, dt\otimes dt + e^{2\xi(t,x)}\, h_{ij}(t) \, dx^i \otimes dx^j \, .
\end{equation}
In particular, $h_{ij}(t)$ can be reduced to $a(t)^2 \, \delta_{ij}$ only in the shear-free case, that is, when the expansion tensor $\theta_i^j=e^{\phi}\,(2\dot{\xi}\,\delta_i^j +\, h^{jk}\dot{h}_{ki})/2$ is everywhere proportional to the identity. 
\end{cor}

\subsection{The Einstein's Field Equations}\label{Sec:GeneralAspects_EinsteinEquations}
We now describe the Einstein's field equations (EFE) for space-flat spacetimes in the adapted coordinates used in equation (\ref{Eq:Metric_Shearfree}). We assume that the vector field $\normal$ is shear-free and the matter content is represented as a fluid idealization co-moving along its flux, in which case the energy-momentum tensor turns into \cite{ellis_mac_marteens,Hwk_Ellis}:
\begin{equation}
\mathbf{T} = \rho \, \normal^\flat \otimes \normal^\flat + p\, \submetric + \normal^\flat\otimes \mathbf{q}^\flat + \mathbf{q}^\flat\otimes\normal^\flat + \Pi,
\end{equation}
where $\rho$ is the energy density, $p$ is  isotropic thermodynamic pressure, $\mathbf{q}$ is the relativistic momentum density and $\Pi$ is the anisotropic pressure.

The $0-0$ component of the EFE provides the generalized Friedmann equation
\begin{equation}\label{Eq:EinsteinFriedmann}
\frac{1}{2}\num{(m-1)(m-2)}\hubble^2 = e^{2\phi}  \rho \, ,
\end{equation}
while the $0-i$ components set the thermodynamic phenomenology relating the relativistic momentum density and the acceleration of $\normal$ through
\begin{equation}\label{Eq:EinsteinFluxq}
q_{i} = \num{(m-2)} \hubble \, \p_i e^{-\phi},
\end{equation}
where we introduce the Hubble parameter $\hubble(t)$ in the usual way
\begin{equation}\label{Eq:HubbleParameter}
\hubble(t) = \frac{1}{a}\, \frac{da}{dt} \, .
\end{equation}
These equations can be used to express the conservation of energy $\nabla_\mu T^\mu_0=0$ as
\begin{equation}\label{Eq:ConservationEnergy}
 \frac{\partial }{\partial t}\left(\, e^{2\phi} \rho \right) =
 \num{(m-1)}\hubble \, \left[ \,-\,  e^{2\phi} \left( \, \rho + p \right) + \num{(m-2)} \left( \frac{e^{\phi } \, \laplacian\,e^{\phi}}{\num{(m-1)}\, a^2}  + \hubble \dot{\phi} \right) \right],
\end{equation}
where $\laplacian$ stands for the Euclidean Laplacian $\laplacian\phi:=\delta^{ij}\p_i\p_j\phi$. Written in terms of $\hubble$, it reads
\begin{equation}\label{Eq:EinsteinDotHubble}
\dot{\hubble} =
- \, \frac{e^{2\phi}}{\num{m-2}} \left( \rho + p \right)
+ \frac{e^{\phi } \, \laplacian\,e^{\phi}}{\num{(m-1)}\, a^2}  + \hubble \dot{\phi} \, .
\end{equation}
It is clear that for $\phi=0$ the formulas (\ref{Eq:EinsteinFriedmann}), (\ref{Eq:ConservationEnergy}) and (\ref{Eq:EinsteinDotHubble}) reduce to the standard equations in a flat RW spacetime.

The last part of the Einstein's equations is the traceless spatial one given explicitly as
\begin{equation}\label{Eq:EinsteinEquationsShear}
\frac{1}{m-1} \, \laplacian e^\phi \, \delta_{ij} - \p_i\p_j e^\phi = e^{\phi}\, \pi_{ij},
\end{equation}
which is in fact an equation of state, in the terms discussed in \cite{mimoso,bss14,BGS}
\begin{equation}\label{Eq:EinsteinEquationsWeyl}
\pi_{ij}=- \frac{(m-2)}{(m-3)} \, E_{ij} \, ,
\end{equation}
where $E_{ij}$ is the electric part of the Weyl tensor. Note that it is a consequence of the non-geodesic and shear-free comoving flow, besides the space-flatness.

It is worth noticing that the generalized Friedmann equation (\ref{Eq:EinsteinFriedmann}) enforces the positivity of $\rho$, except for the static universe $\hubble = 0$, of course. Hence, we can rewrite the conservation law (\ref{Eq:ConservationEnergy}) as an equation involving only $\rho$ and $p$ with the scale factor as a ``time'' variable
\begin{equation}\label{Eq:ConservationEnergy2}
 \frac{\partial \rho}{\partial a} =
 \frac{\num{(m-2)}}{a^3} \, \rho^{1/2} \, \laplacian\,\rho^{-1/2}\, \,
 - \, \frac{\num{(m-1)}}{a}\, \left( \, \rho + p \right) \, .
\end{equation}
As long as the pressure is a known function $p=p(a,x,\rho)$, in principle, we can solve this equation for $\rho(a,x)$. Therefore, the components of energy-momentum tensor and the metric are now functions of the new coordinates $(a,x)$ as we use the identity $e^{2\phi} = \frac{1}{2}\num{(m-1)(m-2)}\hubble^2 \, \rho $ in Eqs.\ (\ref{Eq:Metric_Shearfree}), (\ref{Eq:EinsteinFluxq}) and (\ref{Eq:EinsteinEquationsShear}). So, the relativistic momentum density becomes
\begin{equation}
q_{i} = {\rm sgn}(\hubble) \, \sqrt{\frac{2(m-2)}{m-1}} \, \p_i \rho,
\end{equation}
the anisotropic pressure tensor turns to
\begin{equation}
\pi_{ij} = P_{ij} - \frac{\delta_{ij}}{m-1} \, \delta^{k\ell}\, P_{k\ell} \, ,
\end{equation}
where $P_{ij} = (  2\, \rho \, \partial_i \partial_j\, \rho - 3\,  \p_i \rho\,\p_j \rho)/4 \, \rho^2$ and the metric is
\begin{equation}\label{Eq:Metric_Shearfree_a}
 \metric = - \, \frac{2(m-1)(m-2)}{a^2 \, \rho (a,x)}\,da\otimes da + a^2 \, \delta_{ij} \, dx^i\otimes dx^j \, .
\end{equation}
Note that $\hubble=\hubble(a)$ is a free function independent from the distribution of matter throughout the space and the equation of state used for $p$. Therefore, the problem here is no longer to determine $\hubble(a)$, as in the standard model \cite{ellis_mac_marteens}, but instead, we have to understand the significance of the time variable $t$ that arises when we take $H(a)$ to fit the data. This argumentation will fail only if we return to the homogeneous and isotropic case ($\phi=0$). Indeed, the problem of connecting such solutions with the observations is much more intricate, since the inhomogeneity plays a non-trivial role in it \cite{BGS}.

There are some known solutions of Einstein's equations in the space-flat context. In \cite{coll1_79,coll2_79}, the authors find many characterizations for the solutions in the perfect fluid case. In \cite{Sopuerta_2000}, the co-moving dust solutions are well studied, while in \cite{WolfThomas}, an extensive analysis of the vacuum and general perfect fluid solutions is given. In this manuscript, we keep ourselves as close as possible to the cosmological applications of such spacetimes. Hence, we look at them in a different perspective and take advantage of the natural structure such models have that is suitable to accommodate an inhomogeneous pattern of matter distribution repeated throughout the universe, as we describe in the next section.  

\subsection{The cosmological cells}
Our aim in this manuscript is to set suitable conditions to ensure that our model will be useful as a cosmological one. We shall accomplish it by mimicking the process of observation and analysis often encountered in the description of our universe: there is a scale of length, $L_0$, typically of some hundreds of megaparsecs (Mpc), so that the universe is assumed to be homogeneous from that point on. This means that at each instant of time, by which we mean that one measured by some special free-falling observers, as those in the critical values of $\phi$ \cite{BGS}, we could divide the universe in boxes of length $L_0$ and assume that virtually the same distribution of mass and energy is encountered in each of them. This is essentially what is meant by homogeneity in the standard $\Lambda$-CDM model. Here, we want to move one step further and consider this internal structure from the beginning. As we are going to show, we have a natural mathematical framework for describing it.

Let us assume each flat space section of $\M$ to be a copy of the Euclidean space, $\Sigma_p = \R^{m-1}$. To any discrete subgroup $\Gamma$ of the group of the Euclidean symmetries, $O(m-1)\ltimes \R^{m-1}$, it can be associated a closed set $\cell \subset \R^{m-1}$ that is repeated throughout the space in similar copies as we apply the elements of $\Gamma$ to it. Such a set is called a fundamental domain of $\Gamma$ \cite{vinberg}. If the action is free, the quotient $\Sigma/\Gamma$ is a flat Riemannian manifold. We are interested in the case when $\cell$ is compact of volume $V_0=L_0^{m-1}$. It plays the role of the boxes described in the last paragraph, so that we will refer to it as a \emph{cosmological cell}, that is, a compact fundamental domain such that the quotient $\R^{m-1}/\Gamma$ is a compact flat Riemann manifold.

To illustrate this scheme, in two space dimensions ($m=3$), any compact orientable flat Riemannian manifold is isometric to a torus $\R^2/\Gamma \cong \mathbb{T}^2$, where the discrete subgroup of translations is generated by two linearly independent vectors $\vec{v}_1, \vec{v}_2 \in \R^2$,  $\Gamma = \{ n_1\, \vec{v}_1 + n_2\, \vec{v}_2 \, |\, n_1,n_2 \in \mathbb{Z} \}$. In this case, the vectors $\vec{v}_1$ and $\vec{v}_2$ generate a grid in $\R^2$, each of its compact parts being a cosmological cell of area $L_0^2=||\vec{v}_1 \times \vec{v}_2||$. In three space dimensions, besides the torus generated by three linearly independent vectors, we have five more non-diffeomorphic classes of grids in $\R^3$  generating a compact orientable flat Riemannian manifold of volume $V_0$. In higher dimensions these classes increase in number \cite{Wolf}.

For the sake of simplicity, one might prefer to think of the cosmological cell as the hypercube of side $L_0$, as follows
\begin{equation} \label{Eq:Hypercube}
 \cell = \left\{ (x_1, \ldots, x_{m-1}) \, | \, -\frac{L_0}{2} \le x_i \le \frac{L_0}{2}  \right\} \, .
\end{equation}
Although this is enlightening, and in fact it is essentially all we can have in dimension $m=3$, we should keep in mind that there are many other non-equivalent possibilities in more than two space dimensions.

In our space-flat model, the universe should be considered homogeneous at large scales $L \gg L_0$, where each cosmological cell can be viewed as a point, but completely inhomogeneous as we go to scales of the order or smaller than $L_0$. Therefore, we must look to the field equations with this in mind, adding to them boundary conditions that allow this interpretation. 

\subsection{The boundary and regularity conditions}
\label{SubSec:PeriodicConditions}
The problem of determining the solution of Einstein's field equations is well defined as far as we set the right boundary and regularity conditions. For this, we assume that any spatial function, as $\phi(t,\cdot)$, $\rho(t,\cdot)$ or $p(t,\cdot)$, for each time $t$, to be defined in an open and dense subset $U \subset \R^{m-1}$ and to be $\Gamma$-periodic. This means that, as "$\cdot$" denotes the action of $\Gamma$ in $\R^{m-1}$ and $f$ is such a function, we demand $U$ to be invariant under $\Gamma$, that is, $\Gamma \cdot U = U$, and for every $x \in U$ and $g \in \Gamma$, $f(g \cdot x)=f(x)$. As an example, if $\Gamma$ has as its cosmological cell  the hypercube defined in equation (\ref{Eq:Hypercube}), then we demand that for any  $x \in U$, $i = 1, \ldots , m-1$, we have $(x^1, \ldots , x^i + L_0, \ldots, x^{m-1})\in U$ and
\begin{equation}\label{Eq:RegularPeriodicBoundaryConditionHypercube}
f(x^1, \ldots , x^i + L_0, \ldots, x^{m-1}) = f(x^1, \ldots , x^{m-1}) \, .
\end{equation}

In order to obtain general results concerning existence and uniqueness of solutions to the Einstein's equations, we need to specify the regularity assumptions for the spatial functions. We will accomplish it by working in the Sobolev spaces $\mathcal{H}^N(\R^{m-1})$ and $\mathcal{H}^N(\R^{m-1}/\Gamma)$, with $\R^{m-1}/\Gamma$ a compact manifold (see \cite{Aubin}, Chapter 2). For instance, any spatial function which is $N$ times continuously differentiable in $\R^{m-1}$ ($\mathcal{C}^N$) and $\Gamma$-periodic has all their derivatives up to order $N$ bounded by a constant $C>0$, that is, $|\partial^\alpha f(x)|\le C$ for every  $x \in U$ and $|\alpha|\le N$, where we have used the usual multi-index notation for partial derivatives, so that it is in the space $\mathcal{H}^N(\R^{m-1})$. Furthermore, any function in the quotient manifold differentiable up to the order $N$, say $\tilde{f} \in \mathcal{C}^N(\R^{m-1}/\Gamma)$, gives rise to a unique $\Gamma$-periodic functions $f \in \mathcal{C}^N(\R^{m-1})$, a fact that can be straightforwardly verified from the relation $f(x)=\tilde{f}(\Gamma\cdot x)$. These aspects will be of great importance in the proof of the theorem \ref{Thm:ExistenceUniqueness}. 

\section{On the solutions for simple fluids}\label{Sec:MainTheorems}

\subsection{Existence, uniqueness and early homogeneity}

Let us now turn to the problem of the existence and uniqueness for the solutions of Einstein's equations. In order to settle our hypothesis on the behavior of the cosmological fluid, we must keep in mind that the purpose in this work is to emphasize the role of local inhomogeneity in the global cosmological dynamics. So, if we define the function $\eqstate (t,x)$ as $ p(t,x) = (\eqstate(t,x) -1)\, \rho(t,x)$, we shall expect to find its averaged behavior on large scales ($L\gg L_0$) in the form $\overline{\eqstate}_{\rm eff}=\overline{\eqstate}_{\rm eff}(t)$. We will also assume it at small scales, that is: \textit{the ratio $p/\rho$ is a known function of the scale factor `$a$' only, that is, $\eqstate=\eqstate (a)$}. In this picture, the continuity equation (\ref{Eq:ConservationEnergy}) in terms of $a$ and $\rho$ turns out to be
\begin{equation}\label{Eq:ConservationEnergyRho}
a\, \frac{\p \rho}{\p a} = -(m-1)\, \eqstate(a) \, \rho
+ \frac{(m-2)}{2\, a^2}\, \left(
\frac{3}{2} \left(\frac{|\nabla \rho|}{\rho}\right)^2 - \frac{\laplacian \rho}{\rho}
\right) \, .
\end{equation}
In terms of the function
\begin{equation}\label{Eq:DefinitionFunctionZeta}
\zeta(a) = -(m-1)\, \int_{a_0}^a \, \frac{\eqstate(a')}{a'}\, da' \,
\end{equation}
and the new ``time'' variable
\begin{equation}\label{Eq:DefinitionFunctionTimeS}
s = -\, \frac{m-2}{2}\, \int_{a_0}^a \, \frac{e^{- \zeta(a')}}{a'^3}\, da'
\end{equation}
it turns into
\begin{equation}\label{Eq:ConservationEnergyRho2}
\frac{\p }{\p s}\left(\rho\, e^{-\zeta(s)} \right) =
\frac{\laplacian \rho}{\rho} -
\frac{3}{2} \left(\frac{|\nabla \rho|}{\rho}\right)^2  \, .
\end{equation}
\begin{thm}\label{Thm:ExistenceUniqueness}
Let $\Gamma$ be a discrete subgroup of isometries of the Euclidean space $\R^{m-1}$, such that $\R^{m-1}/\Gamma$ is a compact manifold, and $\cell$ the associated cosmological cell. Fix $a_0 > 0$, $a_{\rm min} = \inf \, \{ a \, | \, 0< a < a_0 \, , \,  0 < s(a) < \infty \}$ and $\rho_0(x)$ a positive and $\Gamma$-periodic function which is $N$ times continuously differentiable on $\R^{m-1}$,  $N>(m+3)/2$. There is a solution $\rho(a,x)$ of the equation (\ref{Eq:ConservationEnergyRho}) satisfying the following conditions:
\begin{enumerate}[(i)]
\item $\rho$ is the unique solution which is continuous in $(a_{\rm min},a_0]\smtimes \R^{m-1}$, smooth ($\mathcal{C}^\infty$) in $(a_{\rm min},a_0)\smtimes \R^{m-1}$, $\rho(a, \cdot)$ is $\Gamma$-periodic in $\R^{m-1}$ for every $a \in (a_{\rm min},a_0]$ and $\rho(a_0,x)=\rho_0(x)$ for every $x \in \R^{m-1}$;
\item For every $x \in  \R^{m-1}$, we have
\begin{equation}\label{Eq:LimitsForRho}
a_{min} < a \le a_0  \quad \Rightarrow \quad 
\rho_0^{\rm min} \, e^{\zeta(a)}\le \rho(a,x) \le \rho_0^{\rm max}\, e^{\zeta(a)} \, ,
\end{equation}
where $\rho_0^{\rm min}$ and $\rho_0^{\rm max}$ are the maximum and minimum values of $\rho_0(x)$, respectively, and $\zeta$ is defined in (\ref{Eq:DefinitionFunctionZeta}). In particular, $\rho$ is everywhere positive: $\rho(a,x) > 0$.
\end{enumerate}
\end{thm}

\proof

It is enough if we prove our theorem of existence and uniqueness of solutions for the equation (\ref{Eq:ConservationEnergyRho}) in the compact Riemannian manifold (without boundary) $\R^{m-1}/\Gamma$. The unique extension of the main theorem to the entire spacetime follows just as explained in the section \ref{SubSec:PeriodicConditions}.

In the first place, if there is a solution which is continuous in the interval $(a_{\rm min},a_0]$, then, by the positivity of $\rho_0$, the energy density can be described by the function $v(a,x)$ through
\begin{equation}
\label{rho_nu}
\rho = v^2 \, e^{\zeta(a)} \ge 0 \, ,
\end{equation}
in an interval $(a,a_0]$. Moreover, uniqueness of $\rho$ follows from the uniqueness of $v$, at least along the maximal interval $(a_1,a_0]$ where $v(a)$ has no zeros in the cell $\cell$, for $v = \sqrt{\rho \, e^{-\zeta(a)}\, }$.
Hence, we rewrite the continuity equation with the ``time'' variable $s$ given by the equation (\ref{Eq:ConservationEnergyRho2}) as the quasi-linear parabolic one:
\begin{equation}\label{Eq:ConservationEnergyU}
\frac{\p v}{\p s} = \nabla \left( \frac{1}{v^2} \, \nabla v \right), \, \qquad  v(0)\equiv v(0,x)=\, \sqrt{\rho_0(x)} \, .
\end{equation}
As a direct consequence of the propositions 8.3 and 9.8 in chapter 15 of the reference \cite{TaylorIII}, where we use the hypothesis $N > (m+3)/2$, there is only one solution $\tilde{v}$ of this initial value problem in $\R^{m-1}/\Gamma$ which is continuous on $[0,\infty)\smtimes \R^{m-1}/\Gamma$ and $\Diff$ on $(0,\infty)\smtimes \R^{m-1}/\Gamma$. Therefore, as we have argued in the section \ref{SubSec:PeriodicConditions}, there is only one extension $v(s,x)$ to $[0,\infty) \smtimes \R^{m-1} \subset \M$, which is continuous on $[0,\infty)\smtimes \R^{m-1}/\Gamma$ and $\Diff$ on $(0,\infty)\smtimes \R^{m-1}/\Gamma$. 
Hence, The existence and uniqueness of the solution $\rho(a,x)$, continuous on $(a_1,a_0]\smtimes \R^{m-1}$ and $\Diff$ on $(a_1,a_0] \smtimes \R^{m-1}$, follows straightforwardly, as well as the positiveness property, that is, $\rho \ge 0$ everywhere. It remains to show that $a_1=a_{\textrm{min}}$. 

We now proceed to show that the maximal interval $[0,s^*)$ where $v(s)$ has no zeros in the cell is in fact $[0,\infty)$, which ensures the uniqueness and positivity of $\rho$ in the interval $(a_{min},a_0]$. First of all, take a point $x_1 \in \cell$ that is a minimum of $v(s_1)$ with $0<s_1<s^*$. If $\laplacian v(s_1,x_1)\ne 0$ then it is positive, since the point in question is a minimum. By continuity, there must be an open set $I \smtimes U \subset (0,s^*)\smtimes \cell$ containing $(s_1,x_1)$ where
$v \, \laplacian v - \num{2} |\nabla v|^2 > 0$, since $v(s_1,x_1)>0$, $\laplacian v(s_1,x_1) >0$ and $\nabla v(s_1,x_1)=0$. Using the equation (\ref{Eq:ConservationEnergyU}), we conclude that $\frac{\p v}{\p s}>0$ in this interval, that is,
\begin{equation}
s \in I\, : \quad s>s_1 \quad \Rightarrow \quad v(s) > v(s_1) \quad \mbox{in $U$}.
\end{equation}
If $\laplacian v(s_1,x_1) = 0$, let $\Sigma_1$ be the connected component containing $(s_1,x_1)$ of the set in $(0,s^*)\smtimes \cell$ defined by the $m$ equations $\nabla v=0$, $\laplacian v = 0$. If $\Sigma_1$ is a point then there will be $(s_2,x_2)$ with $x_2$ a (local) minimum point of $v$ such that $s_2>s_1$ is arbitrarily close to $s_1$ and $\laplacian v(s_2,x_2) >0$, so that the argument above still works. If not, such a point will be in $\Sigma_1$ and there will be a piecewise differentiable curve $(s,x(s)) \in \Sigma_1$ connecting it to $(s_1,x_1)$. For this curve we have, using the equation (\ref{Eq:ConservationEnergyU}) and the fact that  $\nabla v=0$ and $\laplacian v = 0$ along it,
\begin{equation}
 \frac{d}{ds}v(s,x(s))= \frac{\p v}{\p s}(s,x(s)) =0 \, .
\end{equation}
Therefore we conclude that
\begin{equation}
0 < s_1 < s_2 < s^* \quad \Rightarrow \quad v^{\rm min}(s_1) \le v^{\rm min}(s_2) \, ,
\end{equation}
where $v^{\rm min}(s)$ stands for the minimum value of $v(s)$ in $\cell$. Changing $v \to - v$, we readily see that the analogous relation is satisfied by the maximum $v^{\rm max}(s)$ of $v(s)$ in $\cell$:
\begin{equation}
0 < s_1 < s_2 < s^* \quad \Rightarrow \quad v^{\rm max}(s_2) \le v^{\rm max}(s_1) \, .
\end{equation}
So, by continuity of $v$, we conclude that
\begin{equation}
0 < v^{\rm min}(0) \le v(s) \le v^{\rm max}(0) \, ,
\end{equation}
proving the formula (\ref{Eq:LimitsForRho}) and showing that $s^*=\infty$, that is, $\rho$ is unique.
\qed

Theorem \ref{Thm:ExistenceUniqueness} presents a different perspective for the Einstein's equations. Instead of determining the future behavior of the spacetime from its initial conditions, we have found its past history from the ``final'' or today's configuration. This feature is a consequence of picking up the scalar factor as the time variable together with the intrinsic flatness hypothesis. It does not only tells us that our inhomogeneous picture is mathematically consistent, for past existence and uniqueness are granted as we set our current state, but it also allows us to infer some general aspects of such spacetimes, as for instance, the fact that the spatial homogeneity never decreases as the universe expands.   

In order to make sense of our last statement, let us define the maximum, the minimum and the mean energy density in each cell, respectively, as
\begin{equation}
\rho^{\rm max}(a):=\max_{x \in \cell}\, \rho(a,x) 
\quad , \quad 
\rho^{\rm min}(a):=\min_{x \in \cell}\, \rho(a,x) 
\end{equation}
and
\begin{equation}
\overline{\rho}(a):= \frac{1}{Vol_{m-1}(\cell)}\, \int_\cell \rho(a,x)\, d^{m-1}x \, .
\end{equation}
We define the inhomogeneity modulus $\Delta(a)$ as 
\begin{equation}\label{Eq:DefinitionInhomogeneousModulus}
 \Delta(a):= \frac{\rho^{\rm max}(a)-\rho^{\rm min}(a)}{\rho^{\rm max}(a)} \, ,
\end{equation}
while the density contrast, as we interpret $\overline{\rho}(a)$ to be the mean background mass density \cite{Peebles}, is
\begin{equation}
\delta (a,x) = \frac{\rho(a,x)-\overline{\rho}(a)}{\overline{\rho}(a)}  \, .  
\end{equation}
Both $\Delta$ and $\delta$ are measures of the ``degree of homogeneity'' of the spacetime. In the period when $\Delta \ll 1$ or $|\delta| \ll 1$, the spacetime can be considered spatially homogeneous. From the standard model of cosmology, we should expect a highly homogeneous early universe, with the homogeneity growing as the universe expands. Theorem \ref{Thm:ExistenceUniqueness} tells us that our model is consistent with this picture, for it demands the following corollary:
\begin{cor}\label{Cor:HomogeneousSmallerScaleFactor}
Under the hypothesis of Theorem \ref{Thm:ExistenceUniqueness}, setting $\Delta_0=\Delta(a_0)$, we have 
\begin{equation}\label{Eq:InhomogeneousInequalities}
\fl\quad
a \le a_0  \quad \Rightarrow \quad 
\begin{cases}
0 \le \Delta(a) \le \Delta_0 \le 1\\
- \Delta_0 \le - \Delta(a) \le \delta(a,x) \le \frac{\Delta(a)}{1-\Delta(a)} \le \frac{\Delta_0}{1-\Delta_0}.
\end{cases}
\end{equation}
In particular, $\Delta (a)$ do not decrease in an expanding universe.
\end{cor}
\proof
This comes from the inequality (\ref{Eq:LimitsForRho}). First of all, we note that the same inequality holds for $\rho^{\rm max}(a)$, $\rho^{\rm min}(a)$ and $\overline{\rho}(a)$. Hence, we conclude that 
\begin{equation}
\fl\quad
a \le a_0  \quad \Rightarrow \quad 
\frac{\rho^{\rm min}_0}{\rho^{\rm max}_0} \le\frac{\rho^{\rm min}}{\rho^{\rm max}} \le 
\frac{\rho}{\overline{\rho}} 
\le \frac{\rho^{\rm max}}{\rho^{\rm min}}\le \frac{\rho^{\rm max}_0}{\rho^{\rm min}_0} 
\quad {\rm and} \quad
\frac{\rho^{\rm min}_0}{\rho^{\rm max}_0} \le \frac{\rho^{\rm min}(a)}{\rho^{\rm max}(a)} 
\, .
\end{equation}
The inequalities (\ref{Eq:InhomogeneousInequalities}) follow from the fact that $\rho^{\rm min}=(1-\Delta)\,\rho^{\rm max}$ and $\rho=(1+\delta)\,\overline{\rho}$. If $\Delta (a)$ decreases at $a=a_1$, then we could change the "initial" conditions of the theorem \ref{Thm:ExistenceUniqueness} to that of the instant $a=a_1$, thus contradicting the inequalities just obtained. Hence, $\Delta (a)$ cannot decrease with the increasing of the scale factor.  
\qed

The reader should notice that the choice of the equation of state does not determine a priori the Hubble parameter, except when we return to the RW spacetime. However, there are some ways to overcome this situation. For a better understanding of this fact, let us exemplify it by taking any free-falling observer with its proper time, which it imprints to the whole universe as being the cosmological time function $t$. Let us also assume it is initially at rest with respect to the flat space sections, so that, by direct inspection of the geodesic equations and the uniqueness of their solutions, we readily conclude that this observer is placed along the points $(t,x_0)$ where the spatial gradient of $\phi$, and of $\rho$ as well, vanishes. From this point of view, the metric in the formula (\ref{Eq:Metric_Shearfree}) can be set with $\phi(t,x_0)=0$, which looks like a plane RW metric around these points, and the ``Friedmann'' equation for $m=4$ would appear as $3\hubble(t)^2=\rho(t,x_0)$, just like in the flat standard model. We could do that along the points where the energy density has its maximum, thus obtaining a Hubble parameter $\hubble= \hubble^{max}$, or its minimum, with $\hubble= \hubble^{min}$. If we think that ``the Hubble'' parameter satisfies a phenomenological relation $\hubble=\hubble(a)$, just as in the standard $\Lambda$CDM model, we could be neglecting a multiplicative term of order $\hubble^{min}/\hubble^{max}=\sqrt{1-\Delta}$ in the determination of the density parameters $\Omega$'s. This shows us that the choice of the ``cosmological'' observers, the cosmic time $t$ and any comparison with the $\Lambda$CDM model should be taken with care. We leave this task to another investigation, which is beyond the scope of this manuscript.

\subsection{Some exact solutions}
According to the fourth case discussed in section 6.1.2 of the reference \cite{polyanin}, it is straightforward the comparison with our equation (\ref{Eq:ConservationEnergyU}). Thus, one can find implicit solutions in a traveling-wave form by taking the {\em ansatz}
\begin{equation}
\label{ansatz}
|\vec k|^2\int_{v_{{0}}}^{v \left(s,\vec x \right) }\! \frac{{\rm d}u}{u^2(\lambda u+c_1)}=\lambda s+\vec{k} \cdot \vec{x},
\end{equation}
where $\vec k\in\R^{m-1}$, $\lambda\in\R$, $v_0$ and $c_1\neq0$ are arbitrary constants. It is direct to verify that this expression satisfies the equation (\ref{Eq:ConservationEnergyU}). Besides, this integral can be particularly solved, yielding
\begin{equation}
\frac{1}{v_0} - \frac{1}{v} + \frac{\lambda}{c_1}\ln\left(\frac{\frac{\lambda}{c_1}+\frac{1}{v}}{\frac{\lambda}{c_1}+\frac{1}{v_0}}\right) = \frac{c_1(\lambda s+\vec{k} \cdot \vec{x})}{|\vec k|^2}.
\end{equation}
We can isolate $v(s,\vec x)$ in this equation and find the energy density, as follows
\begin{equation}
\label{explicit_nu}
\fl\quad
\rho(s,x,y)=e^{\zeta(s)}\left \{\frac{\lambda}{c_1}+\frac{\lambda}{c_1}\, W\left[-\left(\frac{c_1}{\lambda v_0} + 1\right)\exp\left(\frac{c_1^2 (\vec{k} \cdot \vec{x}+\lambda s)}{\lambda |\vec k|^2}-\frac{c_1}{\lambda v_0}-1\right)\right]\right \}^{-2},
\end{equation}
being $W(X)$ the Lambert W-function (for details see the reference \cite{olver}). As we are dealing only with real arguments for the Lambert W-function, there are only two branches to be considered: $W_0(X)$ for $X>-1/e$ and $W_{-1}(X)$ for $-1/e \leq X < 0$. Note that the equation (\ref{explicit_nu}) is invariant under the transformations $\vec k\rightarrow \beta\vec k$, $\lambda\rightarrow \beta\lambda$ and $c_1\rightarrow \beta c_1$, where $\beta\ne 0$ is a constant, letting  $\nu_0$ unchanged. It means that the profile responsible for the matter distribution inside the cosmological cell may have the same shape for different scales in terms of the wave parameters $\vec k$ and $\lambda$ by choosing $\beta$ adequately.

\begin{figure}[ht]
\centering
    \begin{subfigure}[b]{0.46\textwidth}
    \centering
\includegraphics[width=\textwidth]{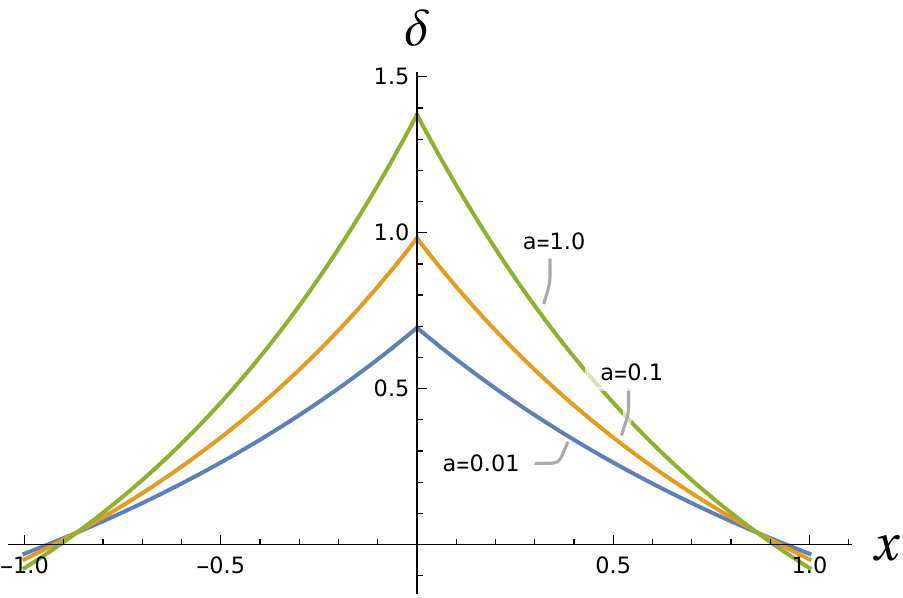}
\caption{The density contrast as function of $x$ (with $y=0$) for different values of $a$.}
\label{delta_x}
\end{subfigure}
\hspace{.3cm}
\begin{subfigure}[b]{0.46\textwidth}
\centering
\includegraphics[width=\textwidth]{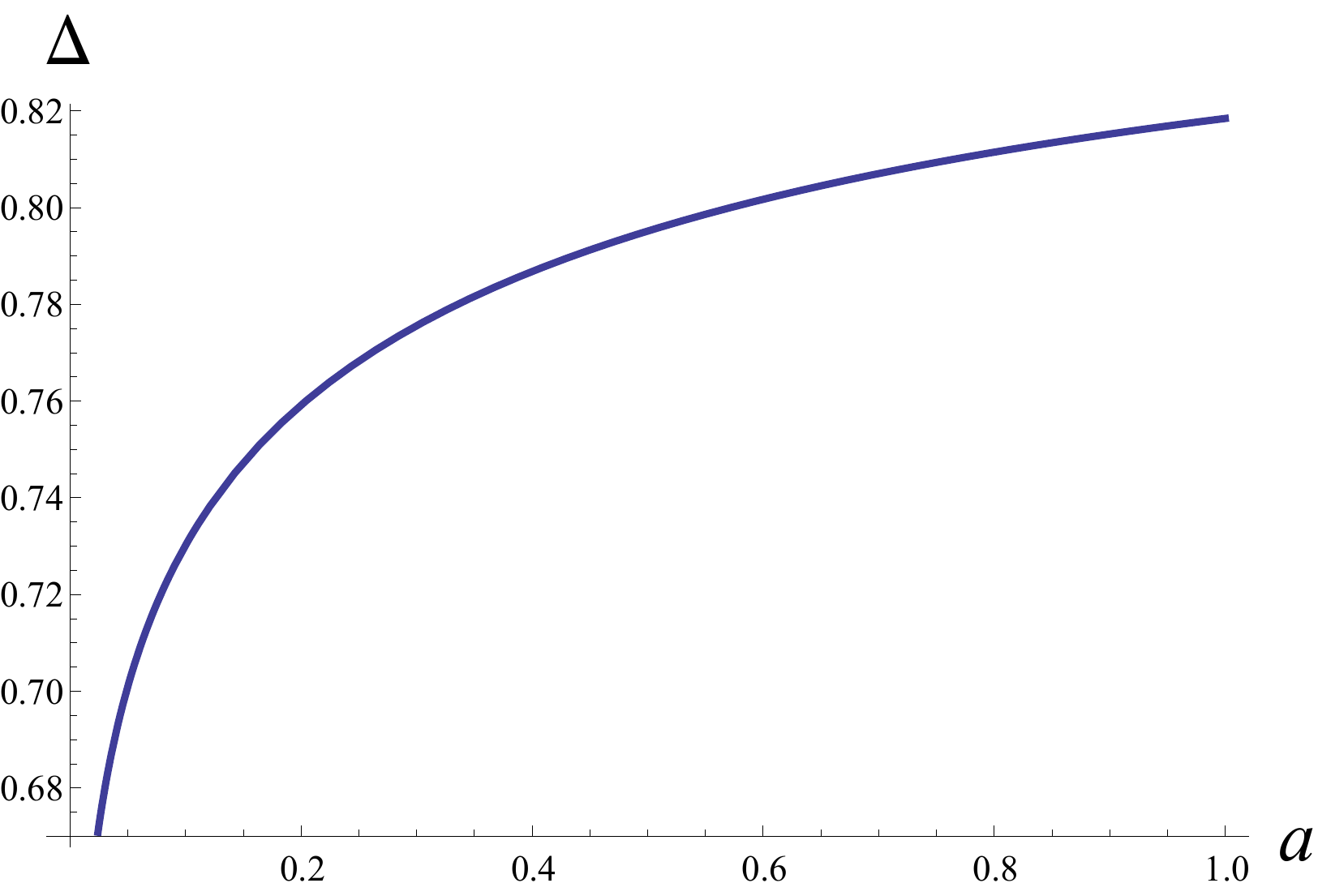}
\caption{The inhomogeneity modulus as function of $a$.}
\label{Delta_a}
\end{subfigure}
\caption{The spatial distribution and time evolution of the inhomogeneity parameters of our model.}
    \label{inhom_par}
\end{figure}

For the sake of illustration, we study a dust-dominated ($\eqstate(a)=1$) cosmological scenario for $m=3$. In this case, the auxiliary functions $\zeta(a)$ and $s(a)$ can be found straightforwardly, reducing to
\begin{equation}
\zeta(a)=2\ln\left(\frac{a_0}{a}\right),\quad\mbox{and}\quad s(a)=\frac{1}{2a_0^2}\ln\left(\frac{a_0}{a}\right),
\end{equation}
\begin{figure}[ht]
    \begin{subfigure}[b]{0.45\textwidth}
             \includegraphics[width=\textwidth]{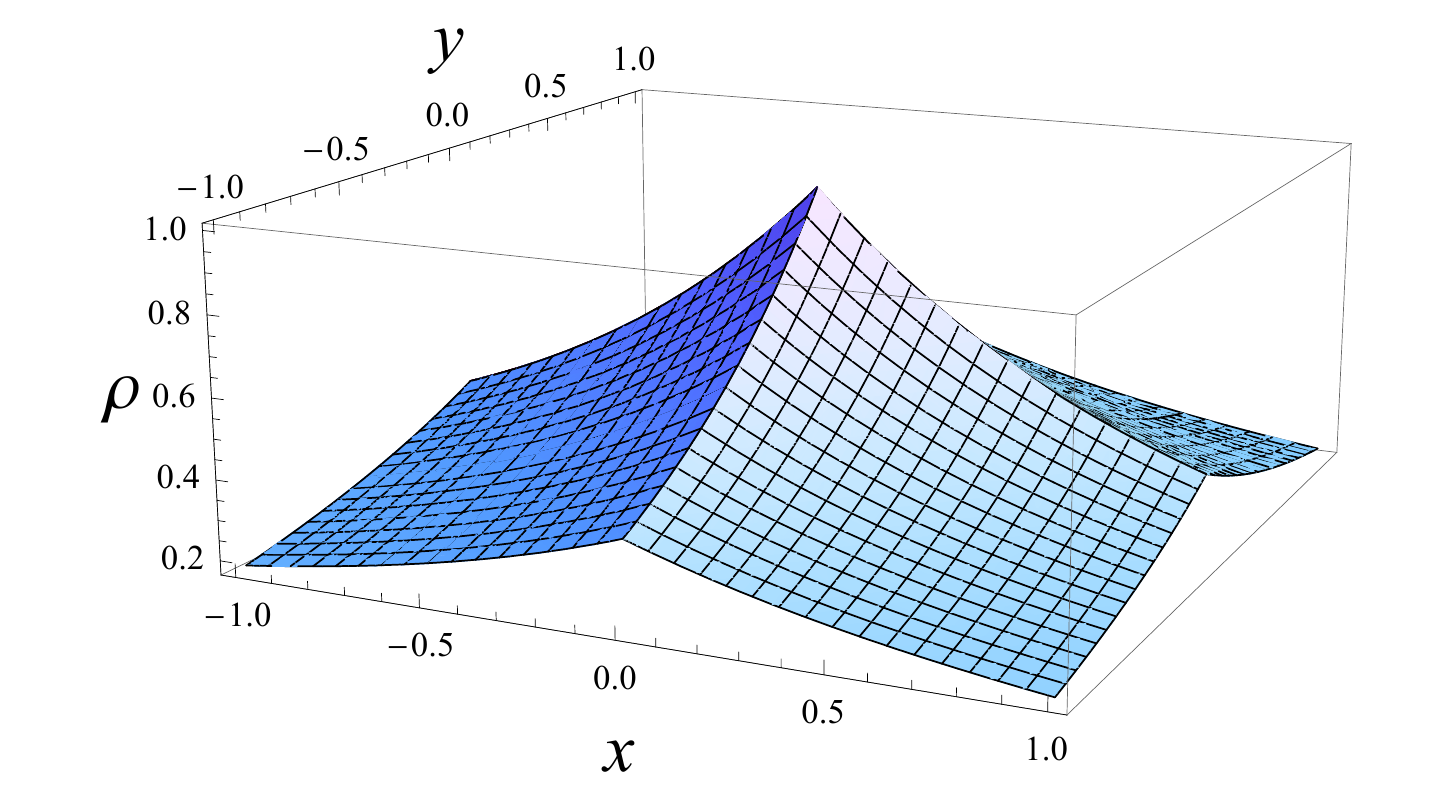}
         \caption{Cosmological cell}
         \label{cosmo_cel}
     \end{subfigure}
     \begin{subfigure}[b]{0.45\textwidth}
           \includegraphics[width=\textwidth]{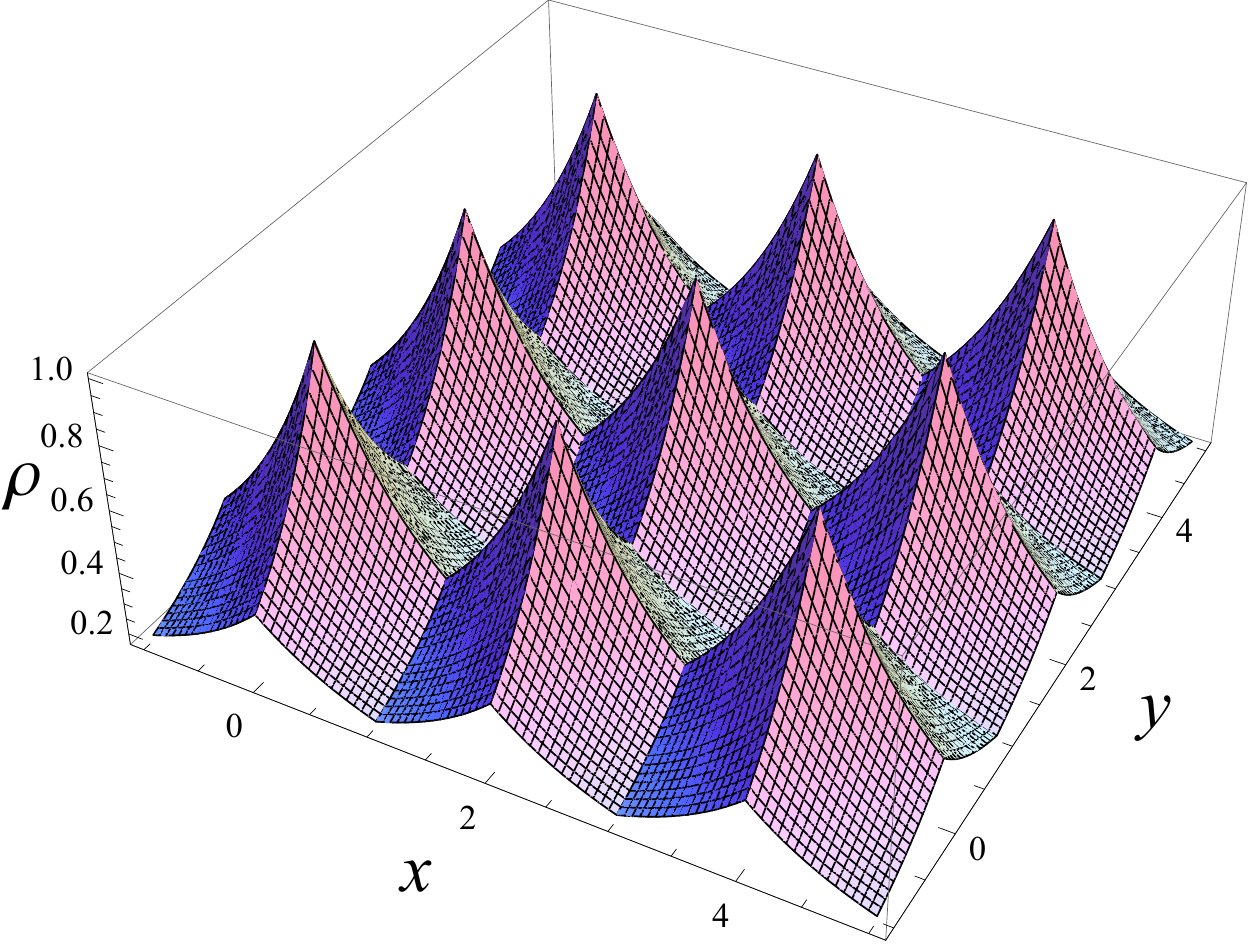}
         \caption{Cosmological lattice}
         \label{cosmo_lat}
     \end{subfigure}
\caption{ Plot of $\rho(1,x,y)$. The cosmological cell (\ref{cosmo_cel}) is designed by taking four sub-cells of unit side, where each of them is given from the equation (\ref{explicit_rho_2d}) with different values for the parameters: $k_x=k_y=1$ for $x\in[0,1]$ and $y\in[0,1]$, $-k_x=k_y=1$ for $x\in[-1,0]$ and $y\in[0,1]$ and so on, respecting the reflection symmetry of the cell. The other constants remain the same for the sub-cells and they are chosen as $\lambda=1$, $c_{1}=-2$ and $v_0=1$. At the end, they are all joined continuously forming a cosmological cell of side $L_0=2$. The cosmological lattice (\ref{cosmo_lat}) is constructed by gluing continuously the boundaries of the cosmological cells.}
    \label{cosmo_cel_lat}
\end{figure}
where $a_0$ is interpreted as the scale factor today and it is set to unit without loss of generality. Then, we can explicitly write $\rho(a,x,y)$ with the help of equations (\ref{Eq:DefinitionFunctionZeta}), (\ref{Eq:DefinitionFunctionTimeS}), (\ref{rho_nu}) and (\ref{explicit_nu}), as follows
\begin{equation}
\label{explicit_rho_2d}
\fl\quad
\rho(a,x,y)=\frac{\lambda}{a^2\,c_1}\left \{1+ W_0\left[-\left(\frac{c_1}{\lambda v_0}+1\right)a^{-\frac{c_1^2|\,\vec k|^2}{2}}\exp\left(\frac{c_1^2(k_1 x + k_2 y)}{\lambda |\vec k|^2}-\frac{c_1}{\lambda v_0}-1\right)\right]\right \}^{-2}.
\end{equation}
We take the cosmological cell to be a square of side $L_0$ centered at the origin of the coordinate system and we divide it in four equal smaller squares of sides $L_0/2$. Thus, we define the solution in $[0,L_0/2]\times[0,L_0/2]$ with the use of formula (\ref{explicit_rho_2d}) and then reflect it in the remaining three parts, say, $[0,L_0/2]\times[-L_0/2,0]$, $[-L_0/2,0]\times[-L_0/2,0]$ and $[-L_0/2,0]\times[0,L_0/2]$. The whole periodic solution is obtained by gluing them throughout the other cosmological cells. The profile of $\rho(1,x,y)$ is depicted by the figure (\ref{cosmo_cel}). Note that we combine four solutions provided by the equation (\ref{explicit_rho_2d}) using different wave parameters in order to get the peaked profile for the energy density. In the figure (\ref{cosmo_lat}), we depicted the cosmological lattice in the studied case where one can see the succession of matter clumps and voids all over the space. Finally, with the help of the figures (\ref{delta_x}) and (\ref{Delta_a}), the former illustrates the density contrast for a spatial section of the cosmological cell and the latter depicts the evolution of the inhomogeneity modulus according to the scale factor. Both present complementary results predicted by the Corollary (\ref{Cor:HomogeneousSmallerScaleFactor}), namely, the growing of the inhomogeneity inside the cell and its time dependence with respect to the scale factor. As $a$ increases the relative inhomogeneities also increase and they vanish as $a$ goes to zero.

For the particular case in which $c_1=0$, the integral in the equation (\ref{ansatz}) is rather trivial and the energy density can be written in terms of elementary functions, as follows
\begin{equation}
\label{explicit_rho_2d_c1_0}
\rho(a,x,y)=\frac{\nu_0^2\,|\vec k|^2}{a^2 \left|\, |\vec k|^2 + \lambda \nu_0^2[\lambda\ln a - 2(k_1 x + k_2 y)]\right|}.
\end{equation}
\begin{figure}[ht]
    \centering
    \includegraphics[scale=1.1]{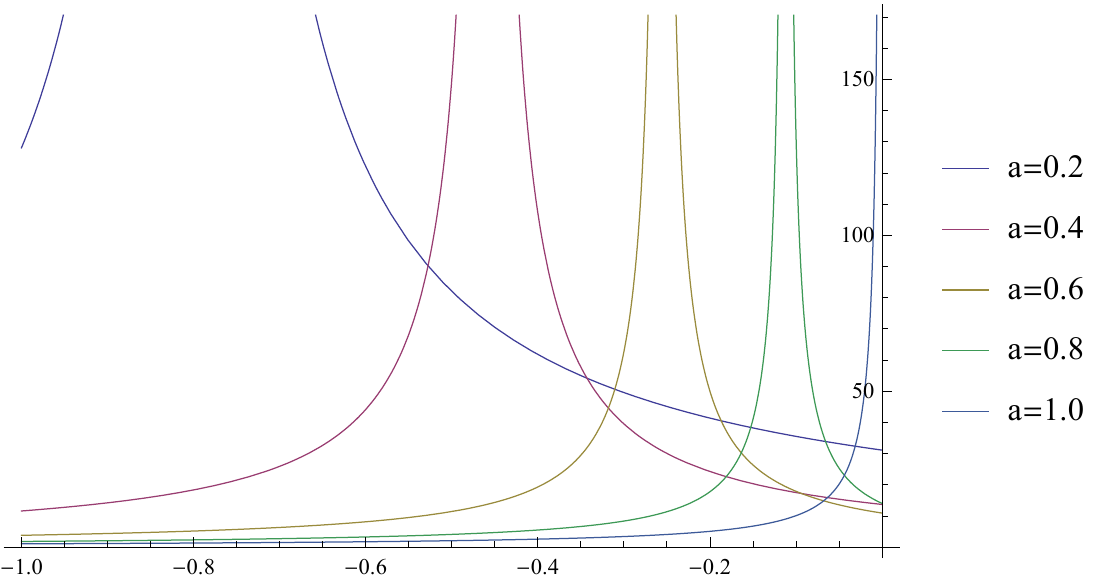}
    \caption{For better illustration, we take the slice $y=1$ of the energy density inside the cosmological cell for different values of $a$, with the parameters $k_x=k_y=\lambda=\nu_0=1$.}
    \label{cosmo_cel_evol_esp}
\end{figure}
Note that now $\rho(a,x,y)$ diverges when the denominator of this equation vanishes. Indeed, from the figure (\ref{cosmo_cel_evol_esp}), it is clear to see the traveling-wave character of this particular solution and the process of concentration of matter as the scale factor evolves. It is worth mentioning that both cases do not satisfy the hypothesis of Theorem  (\ref{Thm:ExistenceUniqueness}), since the energy density is not differentiable all over the cosmological cell.

\section{Concluding Remarks}\label{Sec:Remarks}
From the observations, we know that the universe has inhomogeneites responsible for the emergence of the large scale structures. From the homogeneous and isotropic viewpoint, with a RW background metric, this issue is approached  perturbatively. They are obtained from small quantum fluctuations that start growing when the scale factor is small, evolving to larger scales. According to our results, it is reasonable to expect that models such as ours have ingredients enough to explain qualitatively the same cosmological data in a different way, by taking small perturbations of the fully nonlinear inhomogeneous background, where the matter content is periodically distributed throughout the space. One can find in the literature arguments that these approaches are not equivalent and can lead to difference explanations of the same set of data \cite{Green_2014,Buchert_2015,green2015comments}. Furthermore, it is important to mention that the current standard picture of cosmology is passing through revision, as in the case of the Hubble tension \cite{Riess_2016} or in the explanation of the accelerated expansion \cite{Racz_2017}. Hence, it is fair to exploit other alternatives the General Theory of Relativity offers to us, specially after the period when the inhomogeneities begin to grow significantly. This manuscript is a starting point for one of them, which we think is worth pursuing. 

Starting with an intrinsically flat spacetime filled in with a dissipative fluid, we have established the general mathematical setting for dealing with a cosmological model under periodic boundary conditions, ending up with an inhomogeneous cosmological model defined in a lattice, in which the matter distribution is nontrivial inside the fundamental cells. As we have adapted important theorems from the theory of partial differential equations to our cosmological scenario, we have been able to prove existence and uniqueness theorems of solutions for the Einstein's equations. We also obtained the Corollary \ref{Cor:HomogeneousSmallerScaleFactor}, which ensures their compatibility  with a highly homogeneous phase of the universe for small values of the scale factor. 

We have also found a class of exact solutions for the Einstein's equations, given in terms of the expression (\ref{explicit_nu}) of the energy density, that carries interesting properties, as: (i)  regularity of $\rho$ over the cell for all values of the scale factor; (ii) Spatial homogeneity as $a$ goes to zero ($\Delta \to 0$ as $a \to 0$ for suitable choice of the parameters); (iii) the global picture representing regions of high concentration of matter alternated by voids. Finally, we have studied the particular case of a dust-dominated universe in order to better illustrate the properties of our model, with the cosmological cells and lattice they originate, as well as the asymptotic homogeneous phase in the early universe.

There is a long road to be traced in understanding the space-flat spacetimes with periodic distribution of matter. Some effort to theoretically underpin their observational characteristics have already been made, as the study of the relation luminosity distance $\times$ redshift \cite{BGS}, although a rigorous connection to the observational data is still missing. The many other observational features should be probed in the future, as for instance, the link of such geometric structure with the cosmic microwave background. 

\section*{References}
\bibliography{ref.bib}

\end{document}